\documentclass[prl,superscriptaddress,twocolumn,showpacs,preprintnumbers,amsmath,amssymb]{revtex4}
\usepackage{graphicx}% Include figure files
\usepackage{dcolumn}% Align table columns on decimal point
\usepackage{bm}% bold math

\begin{document}

%\preprint{APS/123-QED}

\title{ $\Lambda N$ correlations from the stopped $K^-$ reaction on ${}^4$He}% Force line breaks with \\

\author{T. Suzuki}
\email{takatosi@riken.jp}
\affiliation{RIKEN Nishina Center, RIKEN, Saitama 351-0198, Japan \\}
\author{H. Bhang}
\affiliation{Department of Physics, Seoul National University, Seoul 151-742, South Korea \\}
\author{J. Chiba}
\affiliation{ Department of Physics, Tokyo University of Science, Chiba 278-8510, Japan \\}
\author{S. Choi}
\affiliation{Department of Physics, Seoul National University, Seoul 151-742, South Korea \\}
\author{Y. Fukuda}
\affiliation{Department of Physics, Tokyo Institute of Technology, Tokyo 151-8551, Japan \\}
\author{T. Hanaki}
\affiliation{ Department of Physics, Tokyo University of Science, Chiba 278-8510, Japan \\}
\author{R. S. Hayano}
\affiliation{Department of Physics, University of Tokyo, Tokyo 113-0033, Japan \\}
\author{\\M. Iio}
\affiliation{RIKEN Nishina Center, RIKEN, Saitama 351-0198, Japan \\}
\author{T. Ishikawa}
\affiliation{Department of Physics, University of Tokyo, Tokyo 113-0033, Japan \\}
\author{ S. Ishimoto}
\affiliation{High Energy Accelerator Research Organization (KEK), Ibaraki 305-0801, Japan \\}
\author{ T. Ishiwatari}
\affiliation{Stefan Meyer Institut f\"{u}r subatomare Physik, A-1090 Vienna, Austria \\}
\author{ K. Itahashi}
\affiliation{RIKEN Nishina Center, RIKEN, Saitama 351-0198, Japan \\}
\author{ M. Iwai}
\affiliation{High Energy Accelerator Research Organization (KEK), Ibaraki 305-0801, Japan \\}
\author{M. Iwasaki}
\affiliation{RIKEN Nishina Center, RIKEN, Saitama 351-0198, Japan \\}
\affiliation{Department of Physics, Tokyo Institute of Technology, Tokyo 151-8551, Japan \\}
\author{\\P. Kienle}
\affiliation{Stefan Meyer Institut f\"{u}r subatomare Physik, A-1090 Vienna, Austria \\}
\affiliation{Physik Department, Technische Universit\"{a}t M\"{u}nchen, D-85748 Garching, Germany \\}
\author{J. H. Kim}
\altaffiliation[Present address: ]{Korea  Research Institute of Standards and Science, Taejon 305-340, South Korea \\}
\affiliation{Department of Physics, Seoul National University, Seoul 151-742, South Korea \\}
\author{Y. Matsuda}
\affiliation{RIKEN Nishina Center, RIKEN, Saitama 351-0198, Japan \\}
\author{H. Ohnishi}
\affiliation{RIKEN Nishina Center, RIKEN, Saitama 351-0198, Japan \\}
\author{S. Okada}
\affiliation{RIKEN Nishina Center, RIKEN, Saitama 351-0198, Japan \\}
\author{H. Outa}
\affiliation{RIKEN Nishina Center, RIKEN, Saitama 351-0198, Japan \\}
\author{M. Sato}
\altaffiliation[Present address: ]{RIKEN Nishina Center, RIKEN, Saitama 351-0198, Japan \\}
\affiliation{Department of Physics, Tokyo Institute of Technology, Tokyo 151-8551, Japan \\}
\author{\\S. Suzuki}
\affiliation{High Energy Accelerator Research Organization (KEK), Ibaraki 305-0801, Japan \\}
\author{D. Tomono}
\affiliation{RIKEN Nishina Center, RIKEN, Saitama 351-0198, Japan \\}
\author{E. Widmann}
\affiliation{Stefan Meyer Institut f\"{u}r subatomare Physik, A-1090 Vienna, Austria \\}
\author{T. Yamazaki}
\affiliation{RIKEN Nishina Center, RIKEN, Saitama 351-0198, Japan \\}
\author{H. Yim}
\affiliation{Department of Physics, Seoul National University, Seoul 151-742, South Korea \\}

\collaboration{KEK-PS E549 collaboration}
\date{\today}% It is always \today, today,

\begin{abstract}
We have investigated correlations of coincident $\Lambda N$ pairs from the stopped $K^-$ reaction on ${}^4$He, and clearly observed $\Lambda p$ and $\Lambda n$ branches of the two-nucleon absorption process in the $\Lambda N$ invariant mass spectra. In addition, non-mesonic reaction channels, which indicate possible exotic signals for the formation of strange multibaryon states, have been identified.
\end{abstract}

\pacs{13.75.Jz, 21.45.+v, 25.80.Nv, 36.10.Gv}% PACS, the Physics and Astronomy
\maketitle

Recently, the existence of strongly bound $\bar{K}$-nuclear states,  in particular, in few baryon systems~\cite{AY1,AY2}, has been intensively discussed. On the experimental side, possible candidates of few-body kaonic systems reported in stopped $K^-$ experiments~\cite{Iwa1,TS1,FINUDA,Piano} were investigated in relation to their non-mesonic decay modes.  
Inclusive and semi-inclusive missing-mass spectroscopy using the ${}^4$He(stopped $K^-$,~$N$) reaction~\cite{Sato,Yim}  showed its limitation in identifying relatively broad states due to the poorly known physical background shape originating from non-mesonic processes. Invariant mass spectroscopy via $\Lambda p$ and $\Lambda d$ correlations in stopped $K^-$ reactions~\cite{FINUDA,Piano,TS2} faced ambiguity in the discrimination of the signal from the direct and indirect contributions of the non-mesonic absorption processes, which are also poorly known. Thus a detailed study of the contribution of non-mesonic multinucleon absorption processes of stopped $K^-$ is crucial for the identification of strange multibaryonic states, especially for those with relatively broad widths. From the theoretical point of view, these processes are expected to be the primary source of the imaginary part of the $\bar{K}$-nuclear potential in the deeply-bound energy region where $\bar{K}N\rightarrow Y\pi$ channels are suppressed, thus dominating the widths of possible bound states. Hence, the dynamics of the absorption processes play an important role in the decay mechanism of possible $\bar{K}$-nuclei.
The presently available information is from bubble chamber experiments which are mainly confined to measurements of total non-pionic capture rates of stopped $K^-$~\cite{Katz}, and the dynamical nature or even their existence are still speculative, except for those on deuterium~\cite{Deuterium}. 
Therefore, comprehensive studies of non-mesonic reactions are indispensable to clarify whether or not the two-nucleon absorption (2NA) exist as well-separable processes and to see whether or not any signatures for multibaryonic states can be clearly identified.

The dynamics of the multinucleon absorption processes are most directly investigated by means of a coincidence measurement of back-to-back-correlated $YN$ pairs from stopped $K^-$ reactions, as
 \begin{eqnarray}
 K^- ``NN"(NN) &\rightarrow& Y N(\tilde{N} \tilde{N}) :\textrm{2NA},\\
 K^- ``NNN"(N)& \rightarrow& Y N\tilde{N} (\tilde{N}) :\textrm{3NA},
 %K^- ``NNNN" &\rightarrow& Y N \tilde{N} \tilde{N} \thickspace \thickspace \medspace :\textrm{4NA},
 \end{eqnarray} 
 where $\tilde{X}$ ($X=N$, $d$, $\pi$, or $\gamma$) denotes an undetected particle $X$, and we adopt the terminology and notations introduced in Ref. \cite{TS2} throughout the paper. Furthermore, the possible di/tri-baryonic signals are expected to appear most clearly in the $YN$ spectra, as
 \begin{equation}
(K^- {}^4\textrm{He})_{\textrm{atomic}} \rightarrow {}^2\textrm{S}^{0/+}_{T=\frac{1}{2}/\frac{3}{2}}+\tilde{N}\tilde{N},\atop
{}^2\textrm{S}^{0/+}_{T=\frac{1}{2}/\frac{3}{2}} \rightarrow Y N ,
\end{equation}
and
\begin{equation}\label{tribaryon}
(K^- {}^4\textrm{He})_{\textrm{atomic}} \rightarrow {}^3\textrm{S}^{0/+}_{T=0/1}+N,\atop
{}^3\textrm{S}^{0/+}_{T=0/1} \rightarrow Y \tilde{N} \tilde{N}.
\end{equation}
This Letter presents the successful measurement of $\Lambda N$ pairs from the stopped $K^-$ reaction in $^4$He in the KEK-PS E549 experiment.

\begin{figure}
\includegraphics[scale=0.225]{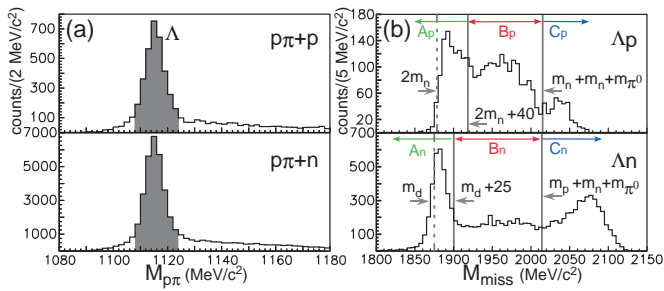}
\caption{\label{fig:pid2} (a) The $M_{p\pi}$ distributions for the $pN\pi$ events. The shaded events are defined as $\Lambda$. (b) The $M_{miss}$ spectra for $\Lambda p$ and $\Lambda n$ events, with three divided zones as defined in the text.}
\end{figure}

Detailed descriptions of the experiment and analysis procedures for proton and neutron channels are given in  Refs.~\cite{Sato} and ~\cite{Yim}, respectively, and the present analysis scheme is covered by Ref. \cite{TS2}. The $p\pi^{\pm}$ invariant mass ($M_{p\pi}$) spectra obtained from $pN$ back-to-back and vertical-$\pi^{\pm}$-coincidence events (see the definition in Ref. \cite{TS2} replacing $dp$ by $pN$ here) show prominent peaks with $\sim$10 MeV/$c^2$ FWHM at the known mass of $\Lambda$ with small combinatorial background. The events in the shaded regions of Fig. \ref{fig:pid2} (a) are defined as $\Lambda$. We identified 3076 and 11003 $\Lambda p$ and $\Lambda n$ pairs, respectively, from $\sim 2.5\times10^8$ stopped $K^-$ in the $^4$He target. The angular region of $-1 \le \cos{\theta_{\Lambda N}} \le -0.6$ was covered, where $\theta_{\Lambda N}$ is the opening angle between the $\Lambda$ and $N$ 3-momenta in the laboratory frame.

\begin{figure*}
\includegraphics[scale=.4]{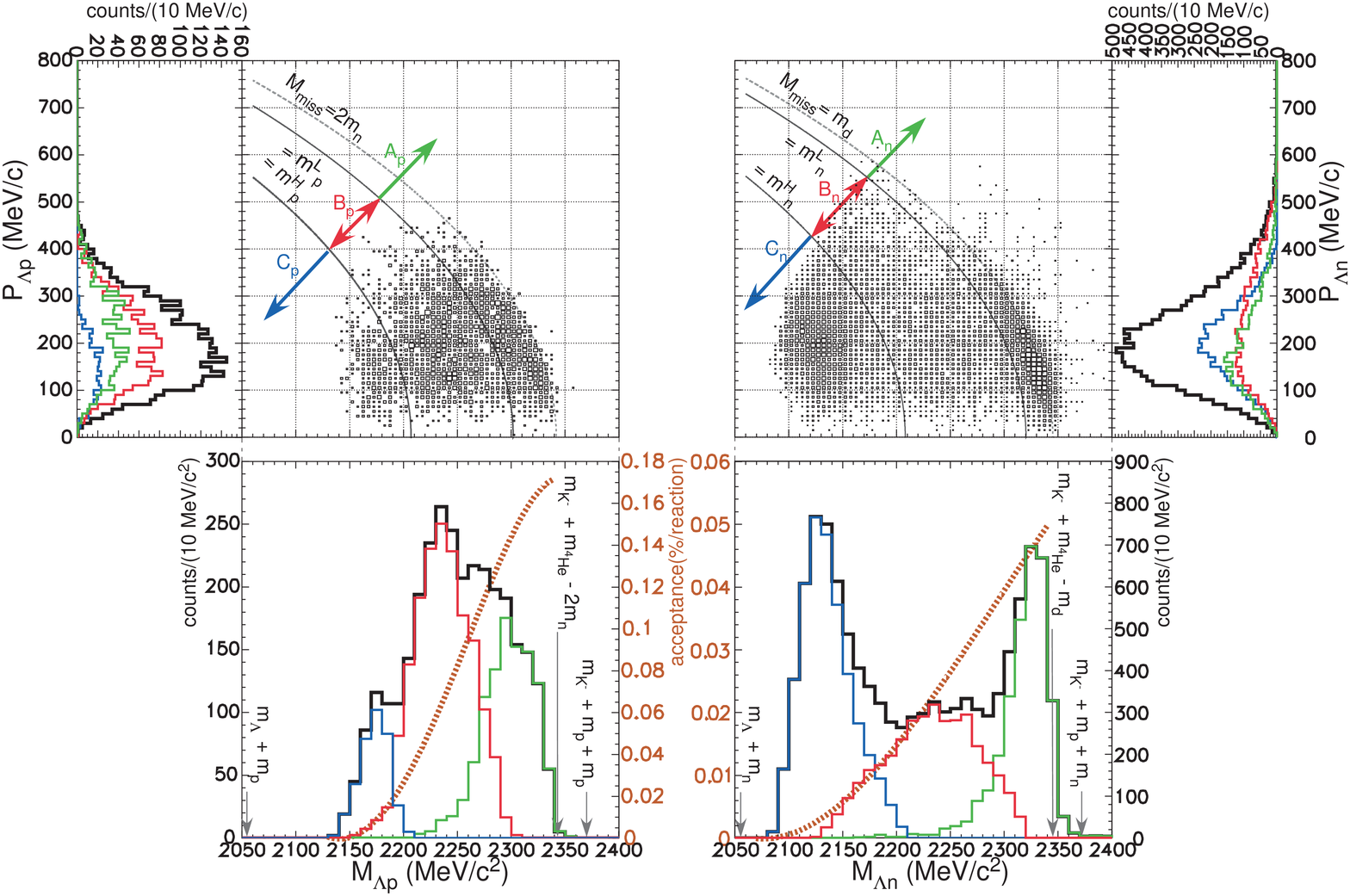}
\caption{\label{fig:inv_totmom} A correlation diagram between $M_{\Lambda N}$(horizontal) and $P_{\Lambda N}$(vertical). In the correlation diagram, $M_{miss}=const$ lines and the event classification zones, A$_N$, B$_N$, and C$_N$ are indicated. The A$_{N}$, B$_{N}$ and C$_N$ components in each of the projections are shown by green, read, and blue lines, respectively. In the projections onto $M_{\Lambda N}$, the evaluated model-dependent acceptance curves are drawn by brown dotted curves, and related mass thresholds are indicated. }
\end{figure*}

 When $\Lambda N$ pairs are detected in the stopped $K^-$ reaction on ${}^4$He, possible final states are $\Lambda N \tilde{N}\tilde{N}$, $\Lambda \tilde{\gamma} N \tilde{N}\tilde{N}$, $\Lambda N \tilde{N}\tilde{N} \tilde{\pi}$, and  $\Lambda \tilde{\gamma} N \tilde{N}\tilde{N} \tilde{\pi}$, where $\Lambda\tilde{ \gamma}$ originates from $\Sigma^0$ and $\tilde{N}\tilde{N}$ includes $\tilde{\textrm{d}}$. These mesonic and non-mesonic final states are separated by reconstructing the missing mass,
\begin{equation}
M_{miss}=\sqrt{(p_{init}-p_{\Lambda}-p_{N})^2},
\end{equation}
where $p_{init}$, $p_{\Lambda}$, and $p_N$ are 4-momenta of the initial state $K^-$+$^4$He at rest,  and the measured ones of $\Lambda$ and $N$, respectively. The distribution of the $M_{miss}$ is shown in Fig. \ref{fig:pid2} (b), for the $\Lambda p$ and $\Lambda n$ cases separately. In both spectra, three prominent components are found. Since $M_{miss}$ represents the total energy of the undetected system in its center of mass frame,  the well-separated structures are considered to imply the contribution of three distinct physics processes. We classify the events into three zones, A$_{p/n}$, B$_{p/n}$ and C$_{p/n}$, as defined by $M_{miss}<m^L_{p/n}$, $m^L_{p/n}<M_{miss}<m^H_{p/n}$, and $m^H_{p/n}<M_{miss}$, respectively, with dividing masses defined as $m^L_p \equiv 2m_n+40.$, $m^L_n \equiv m_d+25.$, $m^H_p \equiv 2m_n+m_{\pi^0}$ and $m^H_n \equiv m_p+m_n+m_{\pi^0}$ in MeV/$c^2$ units. Therefore, the A$_{p/n}$ and B$_{p/n}$ components must be produced in non-mesonic final states, while C$_{p/n}$ can originate from mesonic reactions. 

\begin{figure*}
\includegraphics[scale=.4]{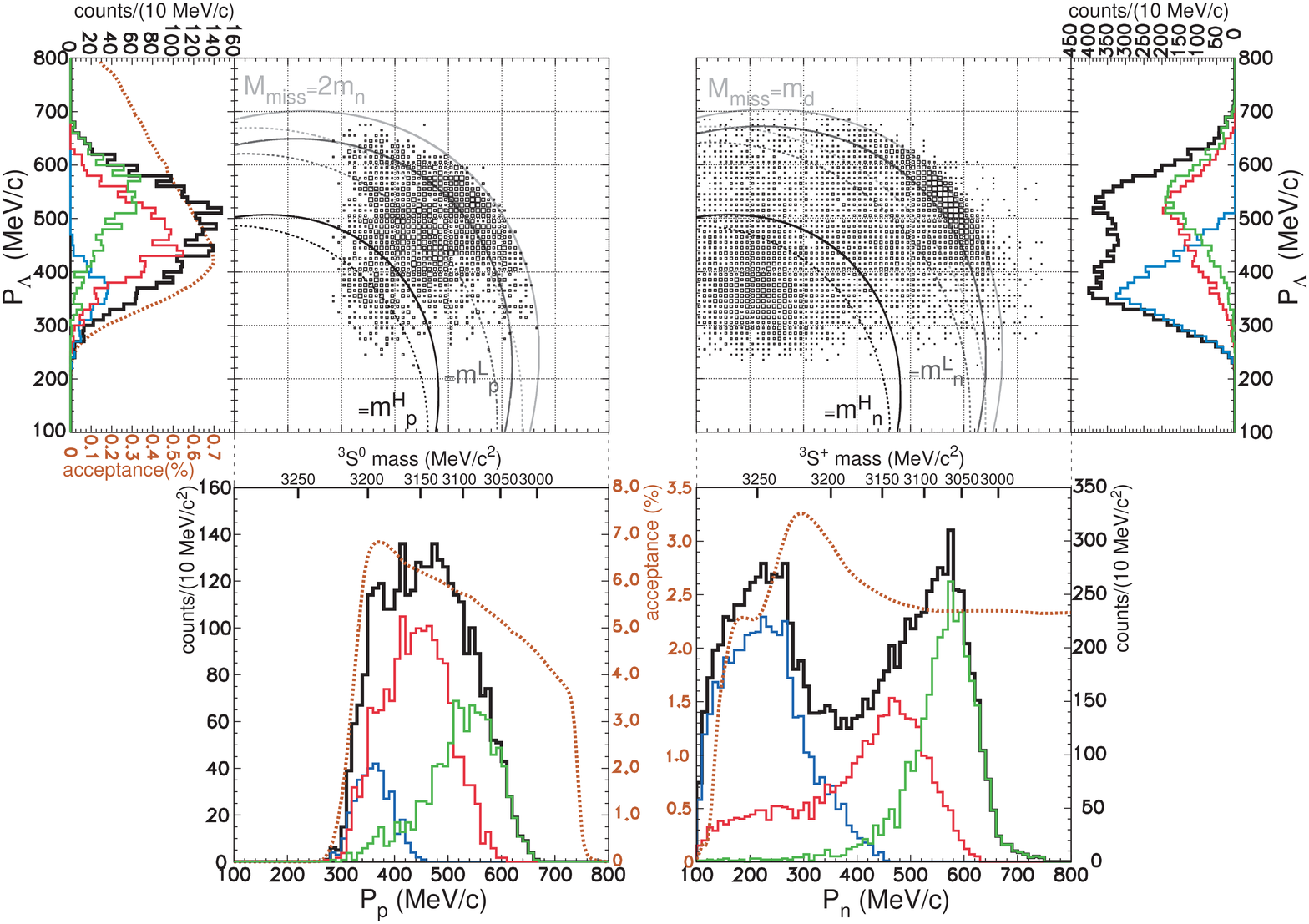}
\caption{\label{fig:mom_corr} A correlation diagram between 3-momenta of $p$ and $n$ (horizontal) and $\Lambda$ (vertical). On the correlations, $M_{miss}=m^H_N$, $=m^L_N$, and $=m_d$ are drawn by black, thick-gray, and thin-gray, with solid and dashed curves for $\cos{\theta_{\Lambda N}}=-1.0$ and $-0.6$, respectively.  On the projections, the A$_{N}$, B$_{N}$, and C$_N$ components are shown by green, red, and blue lines, respectively, and acceptance curves calculated for inclusive $N$ and $\Lambda$ events are drawn by brown dotted curves on $P_p$, $P_n$, and $P_{\Lambda}$ ($\Lambda p$ side) spectra. ${}^4$He(stopped $K^-$, $N$) missing mass (tribaryon mass) scales are shown on the top sides of the $P_N$ spectra.}
\end{figure*}

The correlations between the $\Lambda N$ invariant mass ($M_{\Lambda N}$) and the total 3-momentum ($P_{\Lambda N}$) are shown in Fig. \ref{fig:inv_totmom}, together with the classifications introduced. The figure also shows selected colored and unselected dark spectra projected onto the $M_{\Lambda N}$ and $P_{\Lambda N}$ axes, with acceptance curves for $M_{\Lambda N}$ evaluated by assuming an uniform distribution in the 3-body phase space of ${}^2\textrm{S}^{0/+}+\tilde{N}+\tilde{N}$, and isotropic $\Lambda N$ decay of ${}^2\textrm{S}^{0/+}$ in its center of mass frame, taking the realistic experimental setup and all efficiencies into account. In the spectra shown, acceptance correction is not performed yet, since it is model-dependent, due to which quantitative validity is restricted to a limited $M_{\Lambda N}$ region around the $m_{^4\textrm{He}}+m_{K^-}-2m_N$ thresholds. Without the classification, the $M_{\Lambda p}$ and $M_{\Lambda n}$ spectra show completely different properties. In the unselected $M_{\Lambda p}$ spectrum, no well-resolved peak structure is seen, and the whole strength is broadly distributed below the $m_{^4\textrm{He}}+m_{K^-}-2m_n$ threshold, as was also in the case of the FINUDA experiment \cite{FINUDA}. On the other hand, the unselected $M_{\Lambda n}$ spectrum clearly shows three components, namely, the highest-mass peak located just below the mass threshold $m_{^4\textrm{He}}+m_{K^-}-m_d$, a broad continuum at around $2200\sim2300$ MeV/$c^2$, and the lowest-mass structure toward the threshold of $m_{\Lambda}+m_n$.  

The spectra classified into the A$_{p/n}$, B$_{p/n}$, and C$_{p/n}$ components are interpreted as follows. The A$_{p/n}$ components as asymmetric peaks just below the threshold $m_{^4\textrm{He}}+m_{K^-}-2m_N$ with $M_{miss}$ close to $2m_N$, are interpreted as the $\Lambda N$ branch of the 2NA process, 
\begin{equation}
K^-``NN"(NN) \rightarrow \Lambda N(\tilde{N}\tilde{N}),
\end{equation}
where the undetected residuals are spectators of the reaction. The observed peaks are broader than the $M_{\Lambda N}$ resolutions, $\sim 5$ ($M_{\Lambda p}$) and $\sim 7$ ($M_{\Lambda n}$) MeV/$c^2$ rms in the respective mass regions, reflecting the inherited Fermi motion of the residuals. The spectrum shape of A$_n$ is consistent with the one obtained by a compilation of a $\Lambda d$ spectrum~\cite{ROOSEN}, and the difference between the spectra A$_p$ and A$_n$ in the peak position and width are
well explained by theory~\cite{AY4}. Therefore, this provides clear evidence for a $\Lambda p$ and a $\Lambda n$ branch of the 2NA process, and the existence of the 2NA is now established for the first time as a distinct process which is clearly separated from other non-mesonic reactions. The reaction branching ratio, $BR_{\textrm{A}_p}/BR_{\textrm{A}_n}\sim 0.1$, provides a definite experimental evidence for the dominant contribution of deuteron-like $``pn"$ pairs with spin $S=1$ and isospin $I=0$, and the measured ratio is comparable to or even smaller than the theoretical estimate of $\sim 0.2$~\cite{ONAGA}. The predicted dibaryon $K^-pp$ ($M = 2322$ MeV/$c^2$, $\Gamma_{mesonic} = 61$ MeV/$c^2$)~\cite{AY2} may, if it exists, overlap with this 2NA peak. The formation probability, the decay branching ratio to the $\Lambda p$ channel and the spectrum shape accompanying the reaction chain, should be examined to test its possible contribution to the observed peak. The reaction branching ratios, $BR_{\textrm{A}_p} \sim 0.2\%$ and $BR_{\textrm{A}_n}\sim 2\%$, account for only $\sim20\%$ of the known  $\Lambda NNN$ branching ratio, $11.7 \pm 2.4 \%$ ~\cite{Katz}, and this implies quite unexpectedly that $\sim 80 \%$ of non-mesonic $\Lambda$ final states occupy the yet unassigned components B$_N$. 

The nature of the mysterious components B$_{p/n}$, for which no single step reaction scheme of conventional processes accounts, could include signals of di- or tri-baryonic states, and it should be mentioned that A$_p$+B$_p$ corresponds to the strength which had been attributed to a $K^-pp$ bound state in the FINUDA experiment~\cite{FINUDA}, and thus, further investigation is absolutely needed, as discussed in what follows. On the other hand, the C$_{p/n}$ components with $M_{miss}>2M_N+m_{\pi}$ are considered to be dominated by the quasi-free hyperon production processes, $K^-``N"(NNN)\rightarrow Y\pi (NNN)$, and successive $\Sigma (N) \rightarrow \Lambda N$ conversion, which dominate the stopped $K^-$ reaction \cite{Katz}. 

The correlations between the momenta of the $\Lambda$ and $N$ are shown in Fig. \ref{fig:mom_corr}, together with the overall and selected projections and acceptance curves. There, the A$_{p/n} $ components appear as correlated pairs with the highest momenta. The nucleon momentum ($P_N$) spectra of A$_{p/n}$ components are centered at $\sim 540/580$ MeV/$c$, respectively, with the FWHM$>100$ MeV/$c$ reflecting the Fermi motion, as predicted by PWIA calculation taking the nuclear form factor into account \cite{Iwa2} as well as in Ref. \cite{AY4}. This contradicts the claim for the mono-energetic emission of $N$ and $Y$ \cite{OT} (this aspect is clearly different from the situation observed on ${}^6$Li \cite{FINUDA2}, in which a deuteron with a very small momentum ($\sim50$ MeV/$c$) leads to the reaction, $K^-``d"(^4\textrm{He})\rightarrow \Sigma^- p (^4\textrm{He})$ with monoenergetic protons). 

The B$_{p}$ and B$_{n}$ components show somewhat different behavior. The low-momentum tail of $P_n$, which was detected due to the wide momentum acceptance, is produced by spectators of any reaction. The B$_{p}$ component is centered at $\sim 470$ MeV/$c$ on both $P_p$ and $P_{\Lambda}$. The re-scattering processes proposed to explain the FUNUDA bump~\cite{MORT},
\begin{equation}
 K^- ``NN"(NN) \rightarrow \Lambda \tilde{N}(\tilde{N} \tilde{N}) \atop
 \tilde{N}(\tilde{N}) \rightarrow N' \tilde{N},
\end{equation}
or
\begin{equation}
 K^- ``NN"(NN) \rightarrow \tilde{\Lambda} N (\tilde{N} \tilde{N}) \atop
 \tilde{\Lambda}(\tilde{N}) \rightarrow \Lambda' \tilde{N},
\end{equation}
where $N'$ and $\Lambda'$ are re-scattered particles produced in the primary 2NA processes, account neither for the fact that both momenta are shifted toward lower values by nearly the same amount, nor for the very different B$_p$/A$_p$ and B$_n$/A$_n$ ratios, and thus their contribution should be limited.
The most probable contribution to the B$_{p/n}$ components is a sequence of $\Sigma N$ branches of the 2NA processes and successive $\Sigma\Lambda$ conversion, 
\begin{equation}
K^-``NN"(NN)  \rightarrow  \tilde{\Sigma}N(\tilde{N}\tilde{N}),  \atop
\tilde{\Sigma} (\tilde{N})  \rightarrow  \Lambda \tilde{N}. 
\end{equation}
The observed distributions of masses and momenta are qualitatively consistent with this reaction chain. 
The possible contribution originates from the $\Sigma N$ branches of the 2NA process which are significantly larger compared to the $\Lambda N$ branches reported here. The interpretation would not explain the known $BR_{\Sigma NNN}/BR_{\Lambda NNN}$ fraction \cite{Katz} unless an unexpectedly strong $\Sigma \Lambda$ conversion in the considered momentum region were assumed. Therefore, a quantitative estimation of the cascade processes is vitally important to deduce possible existence of multibaryonic states in the same momentum range. The contribution of the electromagnetic decay, $\Sigma^0 \rightarrow \Lambda \tilde{\gamma}$, can be estimated experimentally by taking the coincidence of $\Lambda N$ and $\gamma$, and accounts for $\sim 30 \%$ of the B$_N$ components.
The $\Lambda NN$ branch of the 3NA process, 
\begin{equation}
K^-``NNN"(N)  \rightarrow  \Lambda N \tilde{N} (\tilde{N}),  
\end{equation}
for which its $\Lambda d$ branch has been observed very recently \cite{TS2}, is another possible source. However, this would require an even larger branching ratio of the 3NA compared to that of the 2NA to account for the entire observed strength.

 One possible exotic interpretation for B$_p$ and B$_n$ is the production of tribaryon states ${}^3$S$^{0}_{T=1}$ and ${}^3$S$^{+}_{T=0/1}$, respectively, and their decay to $\Lambda NN$ or $\Sigma^0 NN$. The scales of the masses corresponding to the momenta of $N$ in the reaction (\ref{tribaryon}) are shown parallel to the momentum scales. If the peaks in the nucleon momentum spectra (red histograms) are interpreted in this way, their masses would be close to 3140 MeV/$c^2$. Another possibility is the productions of the ${^2}$S$_{T=\frac{1}{2}}$ dibaryon and its decay to $\Lambda N$. 
%Among possible dibaryon states, the strongly bound $\bar{K}NN$ states with $``NN"$ nuclear cores with $S=0$, $I=1$ ~\cite{FINUDA, AY2} are hard to account for the observed B$_N$ components, owing to the observed $S=1$, $I=0$ dominance of the 2NA process and small $\Lambda n$ branching ratios even for favored $``pn"_{S=1,I=0}$ pairs as revealed here on $^4$He and reported on deuterium \cite{Deuterium} at zero-energy, and resulting suppression of $\bar{K}NN \rightarrow YN$ decay process which is no more than the 2NA for negative energy $\bar{K}$. 
The possible multibaryonic states enumerated above are considered to have fairly broad spectrum shapes, and hence further investigations are required to allow such interpretations and deduce their binding energies and widths. 

In summary, we have investigated $\Lambda p$ and $\Lambda n$ correlations from the stopped $K^-$ reaction on ${}^4$He, and observed at least two kinds of non-mesonic components in each of the channels. One consists of well-correlated fast $\Lambda p$ and $\Lambda n$ pairs, which are evidence for $\Lambda p$ and $\Lambda n $ branches of the 2NA process in $^4$He, and their dynamical properties have been revealed for the first time. The others are made of slower $\Lambda N$ pairs, which could be interpreted as exotic signals of ${}^2$S$^{0/+}_{T=\frac{1}{2}}$ and/or ${}^3$S$^{0/+}_{T=0/1}$ as well as the cascade of $\Sigma N$ branches of the 2NA and successive $\Sigma \Lambda$ conversion processes. 

We are grateful to the KEK staff members of the beam channel group, accelerator group, and electronics group, for support of the present experiment. We also owe much to T. Taniguchi and M. Sekimoto for their continuous contribution and advices for electronics. This work was supported by KEK, RIKEN, and Grant-in-Aid for Scientific Research (S) 14102005 of the Ministry of Education, Culture, Sports, Science and Technology of Japan.

\bibliography{lncorr}

\end{document}